\documentstyle[aps,prl,floats,epsf]{revtex}
\tighten
\draft
\begin{document}

\twocolumn[\hsize\textwidth\columnwidth\hsize\csname
@twocolumnfalse\endcsname
\title{Technique for Direct eV-Scale Measurements of the Mu and
Tau Neutrino Masses Using Supernova Neutrinos}
 
\author{J.~F. Beacom\thanks{Electronic address:
        {\tt beacom@citnp.caltech.edu}}}
\address{Physics Department 161-33, California
Institute of Technology, Pasadena, California 91125}
\author{R.~N. Boyd\thanks{Electronic address:
        {\tt boyd@mps.ohio-state.edu}}}
\address{Departments of Astronomy and Physics, 
The Ohio State University, Columbus, Ohio 43210}
\author{A. Mezzacappa\thanks{Electronic address:
        {\tt mezz@nova.phy.ornl.gov}}}
\address{Physics Division,
Oak Ridge National Laboratory\thanks{Managed by UT-Battelle, LLC,
for the U.S. Department of Energy under contract DE-AC05-00OR22725.}, 
Oak Ridge, Tennessee 37831}
\date{June 1, 2000}
\maketitle

\begin{abstract}
Early black hole formation in a core-collapse supernova will abruptly
truncate the neutrino fluxes.  The sharp cutoff can be used to make
model-independent time-of-flight neutrino mass tests.  Assuming a
neutrino luminosity of $10^{52}$ erg/s per flavor at cutoff and a
distance of 10 kpc, SuperKamiokande can detect an electron neutrino
mass as small as 1.8 eV, and the proposed OMNIS detector can detect mu
and tau neutrino masses as small as 6 eV.  This {\it Letter} presents
the first technique with direct sensitivity to eV-scale mu and tau
neutrino masses.
\end{abstract}

\vspace{0.25cm}]
\narrowtext


{\bf Introduction:\ } Despite decades of experimental effort, the
values of the neutrino masses remain elusive.  While the laboratory
bound on the electron neutrino mass is about 3 eV~\cite{tritium}, the
laboratory bounds on the mu and tau neutrino masses are much weaker:
170 keV~\cite{Assamagan} and 18 MeV~\cite{Barate}, respectively.  Only
recently have neutrino oscillation experiments found strong evidence
for nonzero {\it differences} of squared neutrino masses.  Once
discovered, the values of the neutrino masses may provide important
clues to physics beyond the Standard Model.  In some scenarios, e.g.,
with the see-saw mechanism~\cite{seesaw}, the mu and tau neutrino
masses are expected to be much larger than the electron neutrino mass.
If they are at the eV scale or greater, the neutrino masses could also
be important cosmologically as a component of the long-sought dark
matter.  It is therefore crucial to devise direct tests of the mu and
tau neutrino masses with sensitivity reaching the eV scale.  While
neutrino mass tests based on cosmological considerations may reach the
eV scale, they are indirect (no neutrinos are detected) and depend
upon the other cosmological parameters being independently
known~\cite{cosmo}.

The best known possibility for directly measuring the mu and tau
neutrino masses is by time-of-flight measurements of supernova
neutrinos, comparing the arrival time of the mu and tau neutrinos to
that of the electron neutrinos.  However, this is complicated by the
long intrinsic duration ($\simeq 10$ s) of the neutrino signal and the
fact that its detailed characteristics are model-dependent.  Beacom
and Vogel have shown that a technique based on the average arrival
times $\langle t \rangle$ is model-independent and is sensitive to
delays as small as $\simeq 0.1$ s~\cite{SNmtau}.  This would allow
detection of mu or tau neutrino masses down to 45 eV in
SuperKamiokande (SK) and 30 eV in the Sudbury Neutrino Observatory
(SNO).  If the mu and tau neutrino masses (strictly speaking, those of
the relevant mass eigenstates) are nearly degenerate, as suggested by
the atmospheric neutrino results~\cite{SKatm}, then the sensitivity
would improve by about $\sqrt{2}$.  Unfortunately, it seems difficult
to improve the results with this technique, since the mass sensitivity
scales with the detector mass $M_D$ as $1/M_D^{1/4}$~\cite{SNmtau}.
To reach the few-eV scale would require detectors $10^4$ times larger,
which seems impossible.

In this {\it Letter}, we discuss a new time-of-flight technique for
measuring neutrino masses that {\it can} reach the eV scale.  This
technique is applicable if the proto-neutron star forms a black hole
early enough to abruptly terminate the neutrino signal.  We state only
our most important results; the details will be discussed at length in
a forthcoming paper~\cite{bigpaper}.


{\bf Expected Neutrino Signal:\ } We consider black hole formation
which occurs soon ($\sim 1$ s) after core collapse (other scenarios
are considered in Ref.~\cite{bigpaper}).  Black hole formation is
triggered by accretion, which drives the proto-neutron star mass above
the maximum stable neutron star mass.  The neutrino signal expected in
this scenario has been studied by Burrows~\cite{Burrows88} and
Mezzacappa and Bruenn~\cite{Mezzacappa}.  In these models, the
neutrino luminosities were fairly constant at more than $10^{52}$
erg/s per flavor until abruptly terminated by black hole formation.
In fact, the transition should have a nonzero duration, of order the
light crossing time $2 R/c \simeq 0.1$ ms, as the proto-neutron star
radius shrinks to that of the final black hole.  During the
transition, the gravitational redshift, originally $\simeq 10\%$,
rapidly diverges, truncating the neutrino signal.  Using a
singularity-avoiding code, Baumgarte et al.~\cite{Baumgarte96b}
studied the transition and found its duration to be 0.5 ms.  Thus, we
can consider the neutrino fluxes to be sharply and simultaneously
terminated.

The results below assume a luminosity $L_{BH} = 10^{52}$ erg/s per
flavor at the cutoff time $t_{BH}$, and a distance $D = 10$ kpc.  We
assume the following temperatures: $T = 3.5$ MeV for $\nu_e$, $T = 5$
MeV for $\bar{\nu}_e$, and $T = 8$ MeV for $\nu_\mu$, $\nu_\tau$,
$\bar{\nu}_\mu$, and $\bar{\nu}_\tau$.  It will be shown that the
necessary quantities can be {\it measured} in a realistic situation.


{\bf Neutrino Mass Effects:\ } At lowest order, a neutrino with mass
$m$ (in eV) and energy $E$ (in MeV) will have an energy-dependent
delay (in s) relative to a massless neutrino in traveling over a
distance D (in 10 kpc):
\begin{equation}
\Delta t(E) = 0.515 \left(\frac{m}{E}\right)^2 D\,.
\label{eq:delay}
\end{equation}
The distance is scaled by the approximate distance to the Galactic
center, though a supernova may be detected from anywhere in the
Galaxy.  For the smallest detectable masses, the delay effects will be
visible only after the sharp cutoff, where no events are otherwise
expected.  Since the delays are very small, the luminosities and
temperatures can be taken as constant over the short interval before
$t_{BH}$.  The event rate for $t > t_{BH}$ is~\cite{bigpaper}:
\begin{equation}
\frac{dN}{dt}(t) = C
\left[\frac{L_{BH}}{10^{51}{\rm erg/s}}\right]
\int_0^{E_{max}} \! \! dE\, f(E)
\left[\frac{\sigma(E)}{10^{-42}{\rm cm^2}}\right],
\label{eq:rateafter}
\end{equation}
where $f(E)$ is the neutrino energy spectrum and $\sigma(E)$ the cross
section.  The upper limit $E_{max}$ on the integral allows only delays
as large as $t - t_{BH}$, i.e.,
\begin{equation}
E_{max} = m \sqrt{\frac{0.515 D}{t - t_{BH}}}\,,
\label{eq:emax}
\end{equation}
where the units are as in Eq.~(\ref{eq:delay}).  Note that the time
and neutrino mass dependence appear only through $E_{max}$.  For $t <
t_{BH}$, $E_{max} \rightarrow \infty$, and the rate is constant.  If
the neutrino energy can be measured, as for some charged-current
reactions, then the event rates for different neutrino energies can
easily be obtained.  For an H$_2$O detector, the constant $C$ is
\begin{equation}
C_{\rm H2O} = (1.74/{\rm s})
\left[\frac{M_D}{1{\rm\ kton}}\right]
\left[\frac{10{\rm\ kpc}}{D}\right]^2
\left[\frac{1{\rm\ MeV}}{\langle E \rangle}\right]\,.
\label{eq:CH2O}
\end{equation}
For a Fermi-Dirac spectrum, $\langle E \rangle = 3.15 T$.  The
constant for a $^{208}$Pb detector can be obtained by scaling by the
relative number of targets/kton, i.e., 18/208.

\begin{figure}[t]
\centerline{\epsfxsize=3.25in \epsfbox{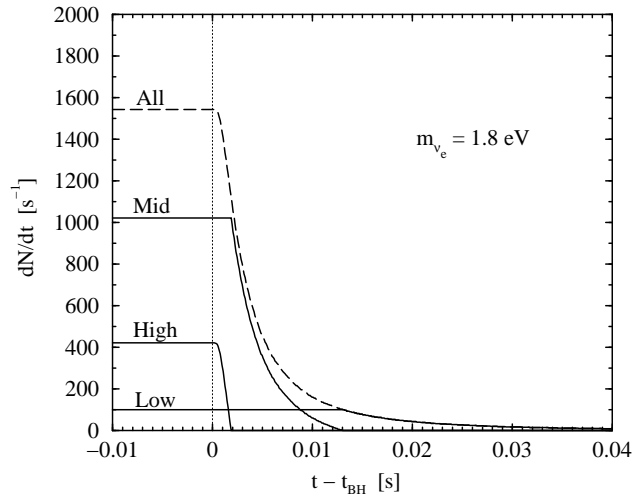}}
\caption{The event rates due to $\bar{\nu}_e + p \rightarrow e^+ + n$
in SK, for different ranges of the neutrino energy: ``Low'' ($0 \le E
\le 11.3$ MeV, contains 2.4 events past the true $t_{BH}$), ``Mid''
($11.3 \le E \le 30$ MeV, 4.8 events), ``High'' ($30 \le E \le \infty$
MeV, 0.5 events), and ``All'' (all energies, 7.7 events).  Note that
only the rate after about $t_{BH}$ is shown, and that the range of $t
- t_{BH}$ is very short.}
\label{fig:nuetest}
\end{figure}

\begin{figure}[t]
\centerline{\epsfxsize=3.25in \epsfbox{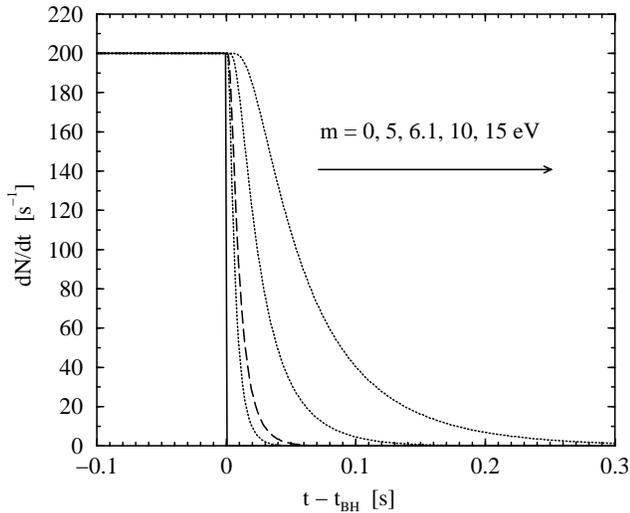}}
\caption{The results for the combined 1-n neutral-current event rate
due to $\nu_\mu$, $\nu_\tau$, $\bar{\nu}_\mu$, and $\bar{\nu}_\tau$ in
OMNIS.  Note that only the rate after about $t_{BH}$ is shown.  Before
$t_{BH}$, other reactions contribute about 20\% of the total neutron
rate; they are not included here, and will have to be statistically
subtracted from the measured rate.  The mu and tau neutrino masses are
assumed degenerate~\protect\cite{SKatm}.  The $m = 0$ case is drawn
with a solid line.  The $m = 6.1$ eV case, with 2.3 events expected in
the tail, is the first case that can be reliably distinguishable from
$m = 0$, and is drawn with a long-dashed line.  The results for other
masses are drawn with dotted lines.}
\label{fig:rateE}
\end{figure}

The expected number of delayed counts after $t_{BH}$ can be calculated
using Eq.~(\ref{eq:rateafter}).  This will be useful when $t_{BH}$ can
be measured independently.  It can be shown~\cite{bigpaper} that this
has the very simple form:
\begin{equation}
N_{del} =
\frac{dN}{dt}(t_{BH}) \times 0.515 \left(\frac{m}{E_c}\right)^2 D\,,
\label{eq:ndel}
\end{equation}
where the event rate is in s$^{-1}$, and the other units are as in
Eq.~(\ref{eq:delay}).  This formula would obviously be true if only a
single energy contributed and the sharp cutoff in the event rate were
rigidly translated by the delay.  But it is remarkable and very
convenient that it is still true even when there is a spectrum of
energies and the event rate develops a decaying tail past the cutoff
(as in Figs.~\ref{fig:nuetest} and \ref{fig:rateE}).  The physical
significance of the ``central'' energy $E_c$ is that it is (to an
excellent approximation) simply the Gamow peak of the falling thermal
spectrum and the rising cross section.  As derived, this is an exact
result.


{\bf Electron Neutrino Mass:\ } We first consider the measurement of
$t_{BH}$ and $m_{\nu_e}$ using the $\bar{\nu}_e + p \rightarrow e^+ +
n$ events in the 32-kton SK detector.  For $T = 5$ MeV, the
thermally-averaged cross section (for the {\it sum} of the two protons
in H$_2$O) is $44. \times 10^{-42}$ cm$^2$~\cite{invbeta}.  The event
rate at or before $t_{BH}$ is thus $\simeq 1500$ s$^{-1}$.  After
$t_{BH}$, the rate is zero if $m_{\nu_e} = 0$ and will develop a tail
if $m_{\nu_e} > 0$.

For a sharp edge, the edge position can be determined with an error
given by the reciprocal of the event rate before the edge, i.e., the
event spacing~\cite{bigpaper,edge}.  If we knew that $m_{\nu_e} = 0$,
then $t_{BH}$ would be determined to $\simeq 1$ ms.  More
realistically, a mass as large as the laboratory bound, $m_{\nu_e}
\lesssim 3$ eV~\cite{tritium}, would cause delays as large as 40 ms,
so that the extracted $t_{BH}$ would be too large.

However, we can simultaneously measure $m_{\nu_e}$ and $t_{BH}$ by
splitting the $\bar{\nu}_e + p \rightarrow e^+ + n$ data into
different ranges of neutrino energy (using $E_\nu \simeq E_{e} + 1.3$
MeV).  These are defined in the caption of Fig.~\ref{fig:nuetest}.
The Low group must be excluded from consideration because these events
have positron total energy less than 10 MeV, and can be confused with
the 5 -- 10 MeV gammas from neutral-current reactions on
$^{16}$O~\cite{oxygen}.  The High group has very little delay and will
thus primarily be sensitive to $t_{BH}$.  Then the Mid group will
determine $m_{\nu_e}$, by counting events delayed past the $t_{BH}$
determined by the High group.

In Fig.~\ref{fig:nuetest}, we show a possible analysis for the case of
$m_{\nu_e} = 1.8$ eV.  In the High group, the number of events in the
tail is $\lesssim 1$, so the cutoff appears sharp and is specified to
within $\simeq 2$ ms.  This uncertainty affects the expected number in
the Mid group by $\simeq 2$ events.  Even so, one can still reliably
see a few delayed counts after the measured $t_{BH}$, enough to
establish a nonzero mass (the statistics are discussed in more detail
below).  A more sophisticated fit would improve our results somewhat,
and we assume a final uncertainty on $t_{BH}$ of about 1 ms.  For a
supernova in which the neutrino fluxes are not truncated by black hole
formation, SK could detect an electron neutrino mass as small as $\sim
3$ eV~\cite{Totani}.


\begin{figure}[t]
\centerline{\epsfxsize=3.25in \epsfbox{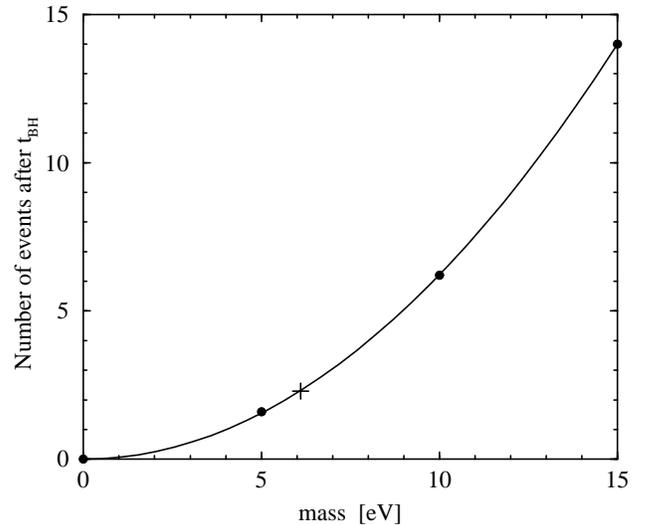}}
\caption{The expected number of delayed counts $N_{del}$ in OMNIS as a
function of the neutrino mass.  The points are obtained by direct
numerical integration of Eq.~(\ref{eq:rateafter}).  The ``+''
indicates the smallest discernible mass at the 90\% CL.  The solid
line is obtained with Eq.~(\ref{eq:ndel}), using $E_c = 40.7$ MeV, the
Gamow peak energy.}
\label{fig:massE}
\end{figure}

{\bf Mu and Tau Neutrino Masses:\ } We consider mu and tau neutrino
detection in OMNIS, a proposed supernova neutrino detector based on
lead and iron~\cite{Boyd}.  Since their energies are below the
charged-current thresholds, supernova mu and tau neutrinos can be
detected only by their neutral-current interactions.  On the other
hand, due to the temperature hierarchy, they will dominate the
neutral-current yields.  In OMNIS, the dominant neutral-current
reaction is the spallation of single neutrons from lead.  The neutrons
could be detected by capture in a gadolinium-doped liquid
scintillator, which yields an 8-MeV gamma cascade in about 0.030 ms
(much smaller than typical mass delays).

For $T = 8$ MeV, the thermally-averaged cross section for the {\it
sum} of $\nu_\mu$ and $\bar{\nu}_\mu$ (or $\nu_\tau$ and $\bar{\nu}_\tau$)
on $^{208}$Pb, including the 1-neutron
spallation probability, is $760 \times 10^{-42}$
cm$^{2}$~\cite{Fuller}.  The cross sections on $^{206}$Pb and
$^{207}$Pb, which together comprise 46\% of natural lead, are expected
to be similar~\cite{bigpaper}.  For a supernova at 10 kpc in which the
neutrino fluxes are not cut off by black hole formation, we assume
that OMNIS will have $\simeq 1000$ 1-neutron neutral-current events
due to $\nu_\mu$, $\nu_\tau$, $\bar{\nu}_\mu$, and $\bar{\nu}_\tau$ on
lead (the events on iron are not included in our calculations).  This
goal could be met with a 2.2 kton lead detector with perfect neutron
detection efficiency.  A realistic design based on 4 kton of lead
and 10 kton of iron, and with about this many events, is described by
Boyd~\cite{Boyd}.

In Fig.~\ref{fig:rateE}, we plot the relevant neutral-current rate for
different values of the neutrino mass, calculated using
Eq.~(\ref{eq:rateafter}).  In Fig.~\ref{fig:massE}, we plot the number
of delayed events $N_{del}$ as a function of the neutrino mass, using
Eq.~(\ref{eq:ndel}) and by direct integration.
Equation~(\ref{eq:ndel}) is remarkable for its simplicity, and also
because it is written in terms of measurable quantities.  The cutoff
time $t_{BH}$ will be measured in SK.  The neutral-current event rate
at or before $t_{BH}$ will be measured in OMNIS, as will $N_{del}$.
The central energy $E_c$ depends on the mu and tau neutrino
temperature, which can be estimated by the neutral-current yields on
different targets~\cite{bigpaper}.  We assume that the distance $D$
can be determined by astronomical means.

Given the measured value of $N_{del}$, Eq.~(\ref{eq:ndel}) can be
immediately solved for the best-fit neutrino mass.  If $N_{del} = 0$
is measured, then the best-fit mass is $m = 0$, and an upper limit can
be placed.  An expectation of 2.3 counts fluctuates down to 0 counts
only 10\% of the time.  Thus, setting $N_{del} = 2.3$, an upper limit
on the mass $m_{lim}$ is obtained.  This is the largest mass, given
the expected Poisson statistics, that could be confused with the
massless case.  For the present case, this is 6.1 eV.

Since the fractional error on $N_{del}$ due to Poisson statistics is
large ($\simeq 1/\sqrt{2.3} \simeq 65\%$), errors on other inputs are
expected to be irrelevant.  The uncertainty on $t_{BH}$ from SK is
assumed to be about 1 ms.  From Fig.~(\ref{fig:rateE}), this
uncertainty can be seen to change the expected number $N_{del}$ by
$\simeq 0.2$ events, which is negligible.  Other possible errors,
e.g., the detector background, the disregarded 0.5 ms tail of the
luminosity, and $\nu_e$ and $\bar{\nu}_e$ events after $t_{BH}$, are
even less important~\cite{bigpaper}.

For a supernova that does not have the sharp cutoff in the rate
characteristic of black hole formation, the model-independent $\langle
t \rangle$ analysis~\cite{SNmtau} yields an $m_{lim}$ that is {\it
independent} of the distance $D$ and scales as
$1/M_D^{1/4}$~\cite{SNmtau}.  For the present case, $m_{lim}$ scales as:
\begin{equation}
m_{lim} \sim E_c \,
\sqrt{\frac{\langle E \rangle \, D}{\sigma_{eff} L_{BH} M_D}}\,,
\label{eq:m2}
\end{equation}
where $\sigma_{eff}$ is the thermally-averaged cross section.  In
terms of absolute sensitivity, these techniques compare as 21 eV and 6
eV, respectively.  These differences are consequences of the sharp
cutoff in the neutrino flux.


{\bf Conclusions:\ } If a black hole forms early in a core-collapse
supernova, then the fluxes of the various flavors of neutrinos will be
abruptly and simultaneously terminated when the neutrinospheres are
enveloped by the event horizon.  For a massive neutrino, the cutoff in
the arrival time will be delayed by $\Delta t \sim (m/E)^2$ relative
to a massless neutrino.

The Galactic core-collapse supernova rate is about 3/century or
higher~\cite{bigpaper}, and the work of Brown and Bethe~\cite{BB}
suggests that black holes are formed about half of the time.  In the
work of Burrows~\cite{Burrows88} and Mezzacappa and
Bruenn~\cite{Mezzacappa}, the neutrino luminosities just before black
hole formation are very high.  These results indicate that there is a
reasonably good chance that such an event could be observed by the
present and proposed supernova neutrino detectors~\cite{bigpaper}.  If
so, there are important practical consequences.

{\it First}, since SK can measure the neutrino energy of the
$\bar{\nu}_e + p \rightarrow e^+ + n$ events, both $t_{BH}$ and
$m_{\nu_e}$ can be measured by the arrival times for different
neutrino energies.  An electron neutrino mass as small as 1.8 eV can
be detected.  {\it Second}, although the mu and tau neutrino energies
are not measured in their neutral-current detection reactions, their
masses can be measured by counting the number of events after
$t_{BH}$.  In the proposed OMNIS detector, a mu and tau neutrino mass
(assumed degenerate~\cite{SKatm}) as small as 6 eV can be detected.
This is the only known direct technique with eV-scale sensitivity for
these masses.  {\it Third}, these results scale with the distance,
luminosity, and detector mass as $\sqrt{D/L_{BH} M_D}$.  This
favorable scaling with the detector mass suggests that it would be
realistic to consider even larger detectors, in order to reach 1 or 2
eV for all three neutrino masses.


\vspace{0.5cm}


J.F.B. was supported by a Sherman Fairchild Fellowship from Caltech.
R.N.B. was supported by NSF grant PHY-9901241.  A.M. is supported at
the Oak Ridge National Laboratory, managed by UT-Battelle, LLC, for
the U.S. Dept. of Energy under contract DE-AC05-00OR22725.  We thank
Felix Boehm, Steve Bruenn, Will Farr, Josh Grindlay, Manoj Kaplinghat,
Gail McLaughlin, Alex Murphy, Yong-Zhong Qian, Petr Vogel, and Jerry
Wasserburg for discussions.



\end{document}